\documentclass{article}
\usepackage{spconfa4,amsmath,graphicx}
\usepackage{amssymb}
\usepackage{booktabs}
\usepackage{tikz}
\usepackage{float}
\usepackage[nolist, nohyperlinks]{acronym}
\usepackage{comment}
\newcommand\norm[1]{\left\lVert#1\right\rVert}

\newcommand{\relu}{\text{ReLu}}
\newcommand{\ldc}{L_\text{DC}}
\newcommand{\lm}{L_\text{Mask}}
\newcommand{\convd}{\text{conv1d}}

\usetikzlibrary{shapes,arrows,positioning,calc}
\tikzstyle{block} = [draw, fill=white, rectangle, 
    minimum height=1.8em, minimum width=2em]
\tikzstyle{mult} = [draw, fill=white, circle, node distance=1cm, path picture={\draw (path picture bounding box.south east) -- (path picture bounding box.north west) (path picture bounding box.south west) -- (path picture bounding box.north east);}]
\tikzstyle{sum} = [draw, fill=white, circle, node distance=1cm, path picture={\draw (path picture bounding box.south) -- (path picture bounding box.north) (path picture bounding box.west) -- (path picture bounding box.east);}]
\tikzstyle{input} = [coordinate]
\tikzstyle{output} = [coordinate]
\tikzstyle{pinstyle} = [pin edge={to-,thin,black}]

\begin{acronym}
\acro{sgm}[SGM]{score-based generative model}
\acro{sota}[SOTA]{state-of-the-art}
\end{acronym}

\title{Robustness of Speech Separation Models for Similar-pitch Speakers}
\name{B. Lay, S. Zaczek, K. Tesch, T. Gerkmann}
\address{Universität Hamburg\\
Signal Processing \\
bunlong.lay@uni-hamburg.de}

\begin{document}
\maketitle

\begin{abstract}
Single-channel speech separation is a crucial task for enhancing speech recognition systems in multi-speaker environments. This paper investigates the robustness of state-of-the-art Neural Network models in scenarios where the pitch differences between speakers are minimal. Building on earlier findings by Ditter and Gerkmann, which identified a significant performance drop for the 2018 Chimera++ under similar-pitch conditions, our study extends the analysis to more recent and sophisticated Neural Network models. Our experiments reveal that modern models have substantially reduced the performance gap for matched training and testing conditions. However, a substantial performance gap persists under mismatched conditions, with models performing well for large pitch differences but showing worse performance if the speakers' pitches are similar. These findings motivate further research into the generalizability of speech separation models to similar-pitch speakers and unseen data.

\end{abstract}
\section{Introduction} \label{sec:intro}

Single-channel speech separation is crucial for robust speech processing in real-world multi-speaker environments. To separate multiple speakers recorded with only one microphone, with the success of Deep Neural Networks it became possible to train approaches that are not speaker-specific and also do not suffer from the permutation problem, which refers to the ambiguity in assigning separated signals to the correct speakers. The first such models were based on Deep Clustering \cite{deep_cluster2016, chimera, chimerapp} and published in 2016 \cite{deep_cluster2016}. As the binary masks that result from clustering may result in artifacts from \cite{deep_cluster2016}, in \cite{chimera} it was proposed to combine Deep Clustering with real-valued mask estimations.
Fully masked-based approaches such as ConvTasnet \cite{luo2019conv} that uses 2D convolutional layers for estimating a mask and permutation invariant training (PIT) \cite{pit}, further improve separation performance. Following this line of research, in \cite{sepformer, mossformer} self-attention from Transformer-based architectures \cite{transformer} was integrated into the mask estimation in 2020 \cite{sepformer}. As a consequence, significant and huge performance improvements on the often used benchmark dataset WSJ0-2mix \cite{deepclustering_wsj0} are made over the first proposed Deep Clustering models \cite{deep_cluster2016, chimera, chimerapp}.

Ditter and Gerkmann showed in \cite{influence_ditter} that Deep Clustering-based models \cite{chimerapp} from 2018 are not robust against mixtures with similar-pitch speakers. More precisely, when such a model is evaluated on a mixture with speakers of similar pitch, then the performance is significantly worse than the performance when tested on a mixture with speakers of very different pitches -- sometimes even worse than the noisy mixture. Since the advancement of more sophisticated Neural Network architectures, we wonder if this performance gap can still be observed.

In the related topic of speech enhancement, it has been shown in \cite{richter2022journal} that \ac{sota} predictive models tend to overfit compared to generative models \cite{richter2022journal, lay202interspeech, lay2024single}. This means predictive models perform much worse on unseen data. Since most \ac{sota} speech separation models are predictive, we wonder if the performance gap is even worse on unseen data.

In this work, we will show that modern \ac{sota} Neural Networks have drastically reduced this performance gap over the years on the benchmark dataset WSJ0-2mix. However, we show that when tested on unseen data the performance gap still remains indicating that further research is necessary to make single-channel speech separation models robust against mixtures of speakers with similar pitches.

\section{Speech Separation}

\subsection{Problem formulation}
The problem of single-channel speech separation can be described as estimating $C$ sources $s_1, \dots, s_C \in \mathbb{R}^{1 \times T}$ from the mixture $x \in \mathbb{R}^{1 \times T}$, where $T$ is the number of samples:

\begin{equation}
    x = \sum_{i=1}^C s_i
\end{equation}
In this work, we consider only mixtures with $C=2$ speakers. In the scientific community mask-based Neural Networks and Deep Clustering-based models are proposed as a solution to the task of speech separation.

\subsection{Mask-based Neural Networks} \label{sec:mask} 
A masker Neural Network is often used to estimate the target sources $s_i$. The Neural Network operates in the time-domain and comprises an encoder-decoder architecture and a masking part. The time-domain mixture $x$ is transformed by a learnable encoder $E(\cdot)$ to a feature space with $F$ many features:
\begin{equation}
    E(x = X \in \mathbb{R}^{F \times T}
\end{equation}
Specifically, we employ $E(x) = \relu(\convd(x))$. Interestingly, it has been shown in \cite{ditter2020} that the learnable encoder of a masker network mimics an STFT-like representation meaning that often $X$ is considered to be in the time-frequency domain. Sequentially, the masker part aims to compute a mask $M_i \in \mathbb{R}^{F \times T}$ for each of the $C$ sources $s_i$. 
The estimate of the target source $s_i$ is then given by $D( X \odot M_i) = \hat{s_i}$. Here $\odot$ is the elementwise multiplication and $D$ is a learnable decoder aiming to invert the encoder by using a transposed convolution layer with the same kernel size and stride as the convolution layer of the encoder. In Section \ref{sec:models} we will discuss different choices for the masking part.

Such a masker model is optimized on the scale-invariant signal-to-noise (SI-SNR) loss in combination with PIT \cite{pit}. The SI-SNR loss \cite{luo2019conv} is defined as
\begin{equation}
    \text{SI-SNR} = 10 \log_{10} \frac{\norm{\alpha \hat{s}}}{\norm{\hat{s} - s}}
\end{equation}
with $\alpha = \frac{\norm{\langle \hat{s}, s\rangle \cdot s}}{\norm{s}}$, where $\norm{\cdot}$ denotes the L2 norm. Many \ac{sota} architectures \cite{luo2019conv, sepformer, mossformer} have followed this scheme over the past years, demonstrating the success of this approach.

\subsection{Deep Clustering}
The main concept behind Deep Clustering \cite{deep_cluster2016} involves utilizing a robust Neural Network to acquire a high-dimensional embedding for each time-frequency unit. The loss function of Deep Clustering results in grouping embeddings of the same speaker closely together in the embedding space, while embeddings from different speakers are further apart. Consequently, straightforward clustering techniques like k-means can be applied to these acquired embeddings to separate sources during testing as is done in \cite{deep_cluster2016}. Specifically, assume that the embedding $v \in \mathbb{R}^{1 \times D}$ corresponds to a pair of time-frequency indices and $y \in \mathbb{R}^{1 \times D}$ is a one-hot encoded vector describing which source is active. With vertically stacked embedding matrix $V \in \mathbb{R}^{TF \times D}$ and vertically stacked one-hot labeled matrix $Y \in \mathbb{R}^{TF \times D}$ the Deep Clustering objective as defined in \cite{deep_cluster2016} is
\begin{equation} \label{eq:deep_clustering}
    \ldc = L(V,Y) = \norm{VV^T - YY^T}_F^2 
\end{equation}
where $ \norm{\cdot}_F $ is the Frobenius norm for matrices.

\section{Research Question} \label{sec:rq}
Isik et. al \cite{isik} reported a significant performance gap between mixtures of same genders versus mixtures of opposite genders for their Deep Clustering approach. This observation motivated the study by Ditter and Gerkmann \cite{influence_ditter} where it has been revealed that there exists a high correlation between the disparity in the pitches of two speakers in the mixture and the performance measured by the Signal-to-Distortion ratio (SDR). %
Specifically, in \cite{influence_ditter} experiments have been conducted on mixtures with two speakers and with Chimera++ \cite{chimerapp}. It has been shown that if the pitch difference $\Delta f_0$ of the two speakers in the mixtures is lower than 60Hz then a steady decline is observed, i.e. when $\Delta f_0$ approaches 0 then it can be expected that the SDR improvement becomes 0. Whereas when $\Delta f_0$ is larger than 60 Hz then averaged SDR improvement is much higher with a much lower standard deviation. 

In this work, we extend the work of Ditter and Gerkmann \cite{influence_ditter}. For the sequel, we refer to \textit{similar-pitch} speakers when the absolute difference of the pitches of the two speakers is less than 60Hz and otherwise, we refer to \textit{different-pitch} speakers. In addition, the \textit{performance gap} of a speech separation model is defined to be the difference between the performance when separating mixtures with different-pitch speakers and the performance with similar-pitch speakers. 

Based on the findings in \cite{influence_ditter}, we wonder if the performance gap still can be observed with more recent \ac{sota} single-channel speech separation models. Therefore, we investigate how the performance of \ac{sota} single-channel speech separation models differs when testing on mixtures with similar-pitch speakers compared to when testing on mixtures with different-pitch speakers. As most \ac{sota} speech separation models are predictive and in \cite{richter2022journal} has been shown that predictive models tend to overfit, we wonder if the performance gap remains virtually the same when tested on unseen data. We will answer these questions in Section \ref{sec:results}.

\section{Analysis Framework} \label{sec:analysis}
We motivated our work and research questions. In this section, we discuss the analysis framework for answering the research questions.

\subsection{Datasets} \label{sec:dataset}
The performance gap w.r.t. difference of pitch $\Delta f_0$ is analyzed on two datasets. These datasets mix two English-speaking speakers with a speaker1-to-speaker2 ratio between $0$ dB and $5$ dB at 8 kHz. \\
\noindent \textbf{WSJ0-2mix:}
The publicly available WSj0-2mix dataset \cite{deepclustering_wsj0} mixes clean speech data from the World Street Journal corpus \cite{wsj0}. It is often used as a benchmark dataset for speech separation with 2 speakers. The dataset contains 30 hours of training data, 7.7 hours for evaluation and almost 5 hours for testing.\\
\noindent \textbf{VCTK-2mix:} 
We use the voice cloning toolkit (VCTK) dataset \cite{yamagishi2019cstr} to mix the VCTK-2mix test set. This test set is used to test the trained models on unseen data. For this, we randomly select two male and two female speakers from this corpus with each having more than 100 utterances. These speakers are then randomly mixed to a length of 4 seconds, resulting in almost 1 hour of data for testing.

\subsection{SOTA single-channel speech separation models} \label{sec:models}
We consider the following speech separation models: \\Chimera++, ConvTasnet, Sepformer and Mossformer, whereas Chimera++ is the only model that is a mask-based and Deep Clustering-based method. In contrast, ConvTasnet, Sepformer and Mossformer are fully mask-based models following the scheme described by Section \ref{sec:mask} which differ only in the masking part. All models were trained on the WSJ0-2mix dataset described in Section \ref{sec:dataset} and achieved \ac{sota} performance on that dataset at the time of their publication. We use online available pre-trained models for Chimera++\footnote{https://github.com/speechLabBcCuny/onssen}, SepFormer\footnote{https://huggingface.co/speechbrain/sepformer-wsj02mix} and MossFormer\footnote{https://github.com/alibabasglab/MossFormer}. \\
\noindent \textbf{Chimera++:}
The Chimera++ model \cite{chimerapp} was introduced as an improvement of \cite{chimera}. This two-headed model outputs feature embeddings and masks for each target source. Both outputs yield two different objectives, namely the Deep Clustering loss $\ldc$ and the mask inference loss $\lm$. The final loss is a combination of both

\begin{equation} \label{eq:dc_mi}
    \alpha \ldc + (1-\alpha) \lm,
\end{equation}
where $1 > \alpha > 0 $ is a hyperparameter. The Chimera++ model \cite{chimerapp} uses a Deep Clustering loss $\ldc$ that is a variation to \eqref{eq:deep_clustering}. This variation discards silent time-frequency bins from the embeddings and applies a weighting matrix to the loss function $\ldc$. For the details we refer to \cite[Eq. (7)]{chimerapp}. In addition, the mask inference loss $\lm$ compares $M_i \odot |X|$ to the truncated phase-sensitive approximation, where $|X|$ denotes the magnitude of the mixture $x$. Moreover, $\lm$ is combined with PIT as it is done for the mask-based approaches. The precise formulation of mask inference loss $\lm$ is given by \cite[Eq. (9)]{chimerapp}. During inference, we discard the Deep Clustering output and utilize only the estimated mask for separation. This model has approximately 9.2M parameters.\\
\noindent \textbf{ConvTasnet:} ConvTasnet \cite[2018]{luo2019conv} is popular benchmark model for single-channel speech separation. The masking part of ConvTasnet employs a cascade of temporal 1D convolutional layers and skip connections. 

We follow the training configuration as in \cite{luo2019conv} to train it on the WSJ0-2mix dataset.
Specifically, we trained ConvTasnet as suggested in \cite{luo2019conv}, i.e. with a learning rate of $10^{-3}$ which is halved after 3 consecutive epochs without improvement on the validation loss and a total of 100 epochs. This model has approximately 5.1 M parameters.

\begin{table*}
\begin{center}
\begin{tabular}{l|c|c||c|c|}
    & \multicolumn{2}{c|}{SI-SDR [dB]} & \multicolumn{2}{|c}{PESQ}\\
    \midrule
         & $\Delta f_0 < 60$Hz & $\Delta f_0 \geq 60$Hz  & $\Delta f_0 < 60$Hz  & $\Delta f_0 \geq 60$Hz\\
\midrule
\midrule
    &\multicolumn{4}{c}{Matched case: \textbf{Tested on WSJ0-2mix}} \\
\midrule
\midrule
    Chimera++ \cite{chimerapp} & 9.34 $\pm$ 3.65 & 11.44 $\pm$ 1.90 & 2.79 $\pm$ 0.45 & 3.15 $\pm$ 0.21 \\
    Conv-TasNet  \cite{luo2019conv} & 12.64 $\pm$ 5.10 & 15.46 $\pm$ 2.33 & 2.80 $\pm$ 0.54 & 3.07 $\pm$ 0.23 \\
    SepFormer \cite{sepformer} & 22.00 $\pm$ 3.51 & 22.94 $\pm$ 1.84 & 3.96 $\pm$ 0.31 & \textbf{4.04} $\pm$ 0.13 \\
    MossFormer \cite{mossformer} & \textbf{22.36} $\pm$ 3.21 & \textbf{23.08} $\pm$ 1.98 & \textbf{3.97} $\pm$ 0.27  & 4.02 $\pm$ 0.12\\
\midrule
\midrule
    &\multicolumn{4}{c}{Mismatched case: \textbf{Tested on VCTK-2mix}} \\
\midrule
\midrule
    Chimera++  & 6.34 $\pm$ 0.51 & 10.08 $\pm$ 4.22 & 2.12 $\pm$ 0.51 & 2.47 $\pm$ 0.44\\
    Conv-TasNet  &  5.33 $\pm$ 4.86 & 9.65 $\pm$ 4.26 & 1.93 $\pm$ 4.26 & 2.26 $\pm$ 0.35 \\
    SepFormer & $13.86 \pm 7.69$ & \textbf{18.68} $\pm$ 5.02 & 3.00 $\pm$ 0.73& 3.42 $\pm$ 0.48 \\
    MossFormer  & \textbf{15.43} $\pm$ 6.81 &   17.89 $\pm$ 5.84 & \textbf{3.24} $\pm$ 0.66  & \textbf{3.47} $\pm$ 0.48\\
\midrule
\end{tabular}
\end{center}
\caption{Performances of speech separation models for the matched and mismatched case. We furthermore distinguish between the similar-pitch case (mixtures where $\Delta f_0 < 60$Hz) and the different-pitch speaker case (mixtures where $\Delta f_0 \geq 60Hz$).} \label{tab:1}
\end{table*} 

\noindent \textbf{SepFormer:} With the rise of Transformers \cite{transformer} for Natural Language Processing, C. Subakan et. al \cite[2021]{sepformer} adopted the mask-based approaches to Transformer architectures. The masking part calculates with the help of self-attention layers short-term dependencies and long-term dependencies along the frames of the encoded mixture. SepFormer has 26M parameters.

\noindent \textbf{MossFormer:} As an improvement to the SepFormer model \cite{sepformer} the Monoraul speech separation TransFormer (MossFormer) \cite{mossformer} has been introduced in 2023. It has been reported that due to the quadratic complexity of the attention layers, SepFormer is still limited to a short context size. The MossFormer model addresses this problem by proposing the gated single-head Transformer with convolution-augmented joint self-attentions. In this work, we use the largest proposed model in \cite{mossformer} with 42.1M parameters.

\subsection{Pitch estimation} \label{sec:freq_estimate}
To estimate the pitch of a speaker in the mixture, we use the PYIN algorithm \cite{pyin} on the clean speech file of that speaker. This algorithm is a probabilistic approach for estimating the pitch $f_0$ in a frame-wise monophonic manner.  
PYIN is composed of two stages. The first stage extracts frame-wise multiple pitch candidates with associated probabilities, and the second stage calculates a smoothed pitch track with a hidden Markov Model. The estimated pitch is then the median of the smoothed pitch track.

\subsection{Metrics} \label{sec:metric}
To evaluate the performance of the speech separation models we use the scale-invariant signal-to-distortion ratio (SI-SDR) as an evaluation metric as defined in \cite{vincent2006performance}. In addition, we also use the Perceptual Evaluation of Speech Quality (PESQ) for objective speech quality testing \cite{rixPerceptualEvaluationSpeech2001}. The PESQ score lies between 1 (poor) and 4.5 (excellent) and we use the narrowband version as we process our audio data in 8kHz.

\begin{figure}[t]
    \centering
    \includegraphics{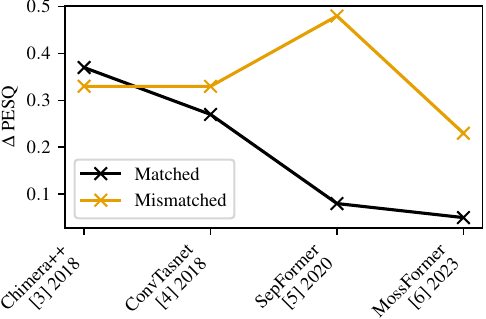}
    \caption{Performance gap between mixtures with similar-pitch speakers and mixtures with different-pitch speakers in the matched testing case (black line) and mismatched case (yellow line).}
    \label{fig:gap}
\end{figure}

\section{Results} \label{sec:results}
We refer to the \textit{matched case} when we test on the WSJ0-2mix test set, as all the models from Section \ref{sec:models} were trained on WSJ0-2mix. The \textit{mismatched case} is the case when we test on the VCTK-2mix test set. 

In Table \ref{tab:1} we observe the SI-SDR and PESQ performance of the Chimera++, ConvTasnet, SepFormer and MossFormer models. For Tab. \ref{tab:1}, we distinguish between the similar-pitch-speaker case (mixtures where $\Delta f_0 < 60$Hz) and the different-pitch speaker case (mixtures where $\Delta f_0 \geq 60$Hz). We observe that the performance in either case (similar-pitch or different-pitch) for the matched case drastically increases from Chimera++ to MossFormer. For instance, Chimera++ achieved only 9.34 dB in SI-SDR for the similar-pitch speaker case, whereas the other models demonstrated subsequent improvements with MossFormer achieving 22.36 dB in SI-SDR. A similar observation can be made for PESQ.

Moreover, the performance gap in the matched case has been sequentially reduced from Chimera++ to the MossFormer mode. This can be seen in Fig. \ref{fig:gap}. Here, we plot the PESQ difference between the different-pitch speaker case to the similar-pitch speaker case for the matched case (black line) and the mismatched case (yellow line). For instance, the value for the matched case for Chimera++ in Fig. \ref{fig:gap} is $3.15 - 2.79 = 0.36$ PESQ difference, whereas the values $3.15$ and $2.79$ are taken from Tab. \ref{tab:1}. We see that MossFormer has a performance gap in terms of PESQ of $4.02 - 3.97 = 0.05$. This indicates that the performance gap is perceptually almost not audible anymore, coinciding with our experience when listening to the separated files.

However, for the mismatched case, we see in Fig. \ref{fig:gap} that the performance gap is hardly reduced with the employment of modern \ac{sota} models (see yellow line). SepFormer has a performance gap of $0.42$ for the mismatched case, which is much larger than the performance gap in the matched case of $0.08$ PESQ. This means that for the different-pitch case SepFormer performs much better than for the similar-pitch case on unseen data. Similar observations can be made for MossFormer and ConvTasnet. Therefore, we conclude that for unseen data, modern \ac{sota} speech separation models still suffer from a lower performance for similar-pitch speakers rather than different-pitch speakers.

\section{Conclusion} \label{sec:conclusion}
In this work, we extended the study of Ditter and Gerkmann \cite{influence_ditter} by investigating if current \ac{sota} speech separation models show robustness against mixtures with similar-pitch speakers. More precisely, we defined the performance gap to be the performance difference when testing a speech separation model on a mixture of different-pitch speakers to similar-pitch speakers. We showed that modern \ac{sota} speech separation models have significantly reduced this performance gap on the benchmark dataset WSJ0-2mix compared to Chimera++ from 2018. Nevertheless, our findings reveal that when tested on unseen data, a performance gap persists, suggesting that \ac{sota} single-channel speech separation models lack robustness when dealing with mixtures of speakers with similar pitch on unseen data.

\newpage
\bibliographystyle{IEEEbib}
\bibliography{main_arxiv}

\end{document}